\documentclass[universe,article,submit,pdftex,moreauthors]{Definitions/mdpi} 
\enlargethispage{\baselineskip}

\firstpage{1} 
\pubvolume{1}
\issuenum{1}
\articlenumber{0}
\pubyear{2022}
\copyrightyear{2022}
\datereceived{} 
\dateaccepted{} 
\datepublished{} 
\hreflink{https://doi.org/}

\pdfoutput=1

\Title{QCD Axion Kinetic Misalignment without Prejudice}
\TitleCitation{QCD Axion Kinetic Misalignment without Prejudice}

\Author{Basabendu Barman$^{1,2}$\orcidA{}, Nicol{\'a}s Bernal$^{1,3}$\orcidB{}, Nicklas Ramberg$^{4}$\orcidC{} and Luca Visinelli$^{5,6*}$\orcidD{}}

\AuthorNames{Basabendu Barman, Nicol{\'a}s Bernal, Nicklas Ramberg and Luca Visinelli}
\AuthorCitation{Barman, B.; Bernal, N.; Ramberg, N.; Visinelli, L.}
\address{%
$^{1}$ \quad Centro de Investigaciones, Universidad Antonio Nari\~{n}o, Carrera 3 este \# 47A-15, 111321 Bogot{\'a}, Colombia\\
$^{2}$ \quad Institute of Theoretical Physics, Faculty of Physics, University of Warsaw, ul. Pasteura 5, 02-093 Warsaw, Poland\\
$^{3}$ \quad New York University Abu Dhabi, PO Box 129188, Saadiyat Island, Abu Dhabi, United Arab Emirates\\
$^{4}$ \quad PRISMA+ Cluster of Excellence \& Mainz Institute for Theoretical Physics, Johannes Gutenberg-Universit{\"a}t Mainz, 55099 Mainz, Germany\\
$^{5}$ \quad Tsung-Dao Lee Institute (TDLI), 520 Shengrong Road, 201210 Shanghai, People's Republic of China\\
$^{6}$ \quad School of Physics and Astronomy, Shanghai Jiao Tong University, 800 Dongchuan Road, 200240 Shanghai, People's Republic of China}
\corres{Corresponding author: luca.visinelli@sjtu.edu.cn}

\abstract{The axion field, the angular direction of the complex scalar field associated with the spontaneous symmetry breaking of the Peccei-Quinn (PQ) symmetry, could have originated with initial non-zero velocity. The presence of a non-zero angular velocity resulting from additional terms in the potential that explicitly break the PQ symmetry has important phenomenological consequences such as a modification of the axion mass with respect to the conventional PQ framework or an explanation for the observed matter-antimatter asymmetry. We elaborate further on the consequences of the ``kinetic misalignment'' mechanism, assuming that axions form the entirety of the dark matter abundance. The kinetic misalignment mechanism possesses a {\it weak} limit in which the axion field starts to oscillate at the same temperature as in the conventional PQ framework, and a {\it strong} limit corresponding to large initial velocities which effectively delay the onset of oscillations. Following a UV-agnostic approach, we show how this scenario impacts the formation of axion miniclusters, and we sketch the details of these substructures along with potential detecting signatures.}


\begin{document}

\section{Introduction}

The QCD axion~\cite{Weinberg:1977ma, Wilczek:1977pj} is a hypothetical pseudo-scalar particle that emerges within the solution to the strong-CP problem proposed by Peccei and Quinn (PQ)~\cite{Peccei:1977hh, Peccei:1977np, Peccei:1977ur}. In the PQ theory, a new global $U(1)_{\rm PQ}$ symmetry is introduced along with a complex scalar field $\Phi$ that is PQ charged. The axion is the angular direction of $\Phi$ after the spontaneous symmetry breaking (SSB) of the $U(1)_{\rm PQ}$ symmetry. QCD anomalies explicitly break the PQ symmetry, reducing it to an approximate global symmetry. More generally, pseudo-scalar particles that couple derivatively to Standard Model (SM) fields are referred to in the literature as axion-like particles (ALPs)~\cite{Arias:2012az}. Axions and ALPs arise in various SM extensions through SSB or from string compactification~\cite{Arvanitaki:2009fg}. Their rich phenomenology allows for numerous experimental approaches which could soon reveal them~\cite{Marsh:2015xka, Irastorza:2018dyq, Arza:2019nta, DiLuzio:2020wdo, Sikivie:2020zpn, Mitridate:2020kly, Co:2021lkc}. Along with this theoretical motivation, the QCD axion is also an excellent particle candidate for explaining the missing dark matter (DM) observed~\cite{Preskill:1982cy, Stecker:1982ws, Dine:1982ah, Abbott:1982af}.

It is currently still unknown when the PQ symmetry breaking occurs concerning other events in cosmology such as inflation~\cite{Turner:1990uz, Lyth:1991ub, Beltran:2006sq, Hertzberg:2008wr, Visinelli:2009zm}. For instance, if the SSB of the PQ symmetry occurs after inflation (i.e. the post-inflationary scenario), it is accompanied by the formation of defects such as networks of strings and walls which do not inflate away and whose relaxation and decay could be a substantial contribution to the axion budget~\cite{Davis:1986xc}. These motivations have pushed for consistently refining the computations regarding the production and evolution of the PQ field in the early Universe and to assess the present relic abundance of the QCD axion. Simulating the formation and decay of the network of strings and domain walls on cosmological relevant scales are mainly subject to errors and uncertainties due to the numerical complexity of the simulation~\cite{Klaer:2017ond, Gorghetto:2018myk, Kawasaki:2018bzv, Vaquero:2018tib, Buschmann:2019icd, Gorghetto:2020qws}. Recent works on axion string simulations indicate that during the non-linear transient regime, the relevance of the axion potential is negligible as long as the gradient terms in the full axion field equation dominates~\cite{Gorghetto:2020qws}. Other recent axion string simulations utilize both the results in Refs.~\cite{Kawasaki:2014sqa, Gorghetto:2020qws} to incorporate the dynamics of the non-linear regime and the contribution from domain walls lead to conflicting results. Note, that domain walls can form even in the absence of a cosmic string network when the field configuration is initially placed close to the hilltop of the axion potential~\cite{Linde:1990yj, Lyth:1991ub} or from the presence of an axion-like particle of mass degenerate with that of the QCD axion at the time of the QCD phase transition~\cite{Daido:2015bva, Daido:2015cba}.

We briefly discuss the issue of domain wall formation in relation to kinetic misalignment, first pointed out in Refs.~\cite{Daido:2015bva, Daido:2015cba} in the context of a two-axions model. In this scenario, one field component acquires a large velocity along a periodic and shallower direction of the potential while being confined along the perpendicular component, until the field settles to some minimum due to cosmic expansion. This is accompanied by a sizable formation of string-free domain walls that are long-lived thus potentially harmful. Such an axion ``roulette’’ mechanism requires the level crossing between two axion species to occur, which is not the case in our model with a single axion. In any case, one method to avoid this potential problem is to consider a small mass for the second axion.

In Ref.~\cite{Buschmann:2021sdq}, the value of the axion mass is expected in the window $m_a \sim \mathcal{O}(40-180)\,\mu$eV, while Ref.~\cite{Hoof:2021jft} suggests a heavier axion mass window, $m_a \sim \mathcal{O}(0.2-80)$~meV, with the discrepancy that can be traced back on different results of the scaling of the spectral index~\cite{Dine:2020pds}. The allowed window of QCD axion mass comprising all of the DM abundance significantly alters on invoking non-standard cosmology~\cite{Visinelli:2009kt, Visinelli:2017imh, Allahverdi:2020bys, Venegas:2021wwm, Arias:2021rer, Ramberg:2019dgi, Arias:2022qjt} or during an era of primordial black hole domination~\cite{Bernal:2021yyb, Bernal:2021bbv, Mazde:2022sdx}.

Nevertheless, despite the level of computation and the extensive literature concerning the dynamics of the axion, there are still many unknown features. For instance, the dynamics of the PQ field in the early Universe could differ from the standard treatment. The initial conditions for the axion field could be set dynamically because of particular choices~\cite{Daido:2017wwb, Takahashi:2019pqf} or the effect of physics beyond the SM~\cite{Co:2018mho, Co:2018phi}. Contrary to the conventional assumptions, the axion field possesses a non-zero initial velocity in the so-called kinetic misalignment (KM) mechanism~\cite{Co:2019jts, Chang:2019tvx}. In this case, the PQ symmetry is broken explicitly not only by QCD anomalies but also by the radial direction of the PQ field. For instance, a global symmetry is generally not a fundamental field property and gets spoiled by quantum effects. Nevertheless the \textit{quality} of the PQ symmetry has to be protected from these effects in order not to jeopardize the solution of the strong-CP problem~\cite{Georgi:1981pu, Dine:1986bg, Giddings:1989bq}, models in which the PQ symmetry is an accidental symmetry explicitly broken by quantum effects have been constructed~\cite{Holman:1992us, Kamionkowski:1992mf, Ghigna:1992iv, Barr:1992qq, Dine:1992vx}. For example, an additional linear term would lead to a wiggly potential and to the fragmentation of the axion field~\cite{Fonseca:2019ypl, Morgante:2021bks}, see also Refs.~\cite{Jaeckel:2016qjp, Berges:2019dgr}, similar to what is expected in models of inflation~\cite{Dolgov:1989us, Traschen:1990sw, Kofman:1994rk, Amin:2010dc}. The additional wiggles in the PQ potential due to the explicit symmetry breaking terms could source the initial non-zero axion rotation and a non-zero PQ charge, which can convert into a matter asymmetry through strong sphaleron processes~\cite{Co:2019wyp}, electroweak sphalerons~\cite{Co:2020jtv}, or in supersymmetric models~\cite{Co:2021qgl}. Observationally, upcoming gravitational wave detectors could be able to distinguish between the standard and kinetic misalignment scenarios~\cite{Gouttenoire:2021wzu, Gouttenoire:2021jhk, Madge:2021abk}.

However, independently on the UV dynamics that sources the non-zero initial axion velocity, one can look for potential observational effects. As the relic abundance of axions becomes immensely altered for specific benchmark values of its initial velocity, we utilize this property in the context of the mass function of axion miniclusters (AMCs). In this work, we explore how the characteristic mass of AMCs is affected by different axion production mechanisms. We first provide a brief review of the production of axion DM in the standard misalignment and the KM mechanisms. Then, we elaborate on the motivation concerning different regimes of the KM mechanism before analyzing its various impacts on the characteristic minicluster mass function as a function of the axion mass. After AMCs form around matter-radiation equality (MRE), the clumping of these structures proceeds to present time. The merging process of the clumping of the matter is a complicated and numerically extensive process typically comprising of N-body simulations~\cite{Eggemeier:2019khm}. Even though numerical simulations are in place to be able to make confident predictions, the intention here is to draw attention to a scenario that could serve as a motivation for N-body simulations in the future. Here we employ a semi-analytic approach from the evolution of a linear density contrast such as the Press-Schechter (PS) formalism~\cite{Zurek:2006sy, Fairbairn:2017sil}. Since KM can be accessibly realized in the pre-inflation scenario, we focus on this case, which consists in invoking non-standard scenarios for the formation of the axion inhomogeneities.

The paper is organized as follows. In Sec.~\ref{sec:production} we briefly review different production mechanisms for axion mentioning the basic equation of motion that governs axion dynamics. We then move on to the discussion of the central theme of this work which is the KM in Sec.~\ref{sec:misalignment}, where we first investigate the weak KM limit in Sec.~\ref{sec:weakkm}, while in Sec.~\ref{sec:strongkm} the strong KM limit is addressed. The impact of KM on the AMC mass and some observational consequences are discussed in Sec.~\ref{sec:minicluster}.
We emphasize that during this work we follow an pragmatic approach, being agnostic about the UV completion of the model that sources the non-zero axion velocity, allowing us to analyze the KM regime without any prejudice. Finally, we summarize our findings in Sec.~\ref{sec:concl}.

\section{Standard scenario}
\label{sec:production}

We consider a SM-singlet complex scalar field $\Phi$, the PQ field, which extends the SM content and which is described by the effective Lagrangian
\begin{equation}
    \label{eq:lagrangian}
    \mathcal{L} = \mathcal{L}_{\rm QCD} + \lvert\partial_\mu \Phi\rvert^2 - V(\Phi) + \mathcal{L}_{\rm int}\,,
\end{equation}
where $\mathcal{L}_{\rm QCD}$ captures all QCD effects in the SM, the PQ field potential responsible for the SSB of $U(1)_{\rm PQ}$ at the energy scale $v_a$ with coupling $\lambda_\Phi$,
\begin{equation}
    \label{eq:mexicanhat}
    V(\Phi) = \frac{\lambda_\Phi^2}{2}\,\left(\lvert\Phi\rvert^2 - \frac{v_a^2}{2}\right)^2,
\end{equation}
and where the term $\mathcal{L}_{\rm int}$ is responsible for the interaction of $\Phi$ with other beyond-SM physics, leading to an effective coupling of the field with gluons and other SM particles. The Lagrangian in Eq.~\eqref{eq:lagrangian} is invariant under the continuous shift symmetry
\begin{equation}
    \label{eq:shiftsymmetry}
    a \to a + \alpha\,v_a\,,
\end{equation}
for a generic value of $\alpha$ that corresponds to a rotation in the complex plane $\Phi \to e^{i\, \alpha}\, \Phi$. After SSB, the complex scalar field can be decomposed in polar coordinates as
\begin{equation}
    \label{eq:complexscalar}
    \Phi = \frac{1}{\sqrt{2}}\, \left(S+v_a\right)\, e^{i\, a/v_a},
\end{equation}
where the angular direction is the axion $a$ and the radial direction is the saxion $S$, such that the saxion vacuum mass is $m_S = \lambda_\Phi\, v_a$.

After SSB, the Lagrangian in Eq.~\eqref{eq:lagrangian} reads
\begin{equation}
    \label{eq:lagrangianSSB}
    \mathcal{L} = \mathcal{L}_{\rm QCD} + \frac{g_s^2}{32\pi}\frac{a}{v_a}\,{\rm Tr}\,G_{\mu\nu}\tilde G^{\mu\nu} + \frac{1}{2}(\partial_\mu a)^2 - V_{\rm QCD}(a)\,,
\end{equation}
where $g_s$ is the QCD gauge coupling of the strong force, and $G_{\mu\nu}$ is the gluon field with dual $\tilde G_{\mu\nu}$. The effective QCD axion potential $V_{\rm QCD}(a)$ arises from the interaction of the axion with QCD instantons around the QCD phase transition~\cite{Gross:1980br} and leads to a mass term for the axion, which would otherwise remain massless in the absence of an explicit breaking of the $U(1)_{\rm PQ}$ symmetry. QCD terms break the continuous symmetry in Eq.~\eqref{eq:shiftsymmetry} explicitly, while leaving a residual $\mathbb{Z}_N$ discrete shift symmetry with $N$ vacua, $a \to a + n\, \pi\, f_a$, with $n$ a natural number and $f_a = v_a/N$ is the axion decay constant. Here, we set $N = 1$. Because of this, the exact form of the axion potential is periodic around the temperature at which the QCD phase transition occurs, $T_\text{QCD} \simeq 150$~MeV. At high temperatures, the shape of the potential is well approximated by a cosine potential meanwhile, for $T \ll T_\text{QCD}$ the shape of the potential is well approximated by its zero temperature prediction which can be computed at the next-to-leading order within perturbation theory~\cite{DiVecchia:1980yfw, GrillidiCortona:2015jxo, Gorghetto:2018ocs}. Here, we adopt the parametrization
\begin{equation}
    \label{eq:axion_potential}
    V_{\rm QCD}(a) = \chi(T)\, \left(1 - \cos\theta\right)\,,
\end{equation}
where $\theta(t) \equiv a(t)/v_a$ is the axion angle and $\chi(T)$ is the QCD topological susceptibility. Much of the recent effort has been devoted to the numerical evaluation of the functional form of $\chi(T)$~\cite{Berkowitz:2015aua, Borsanyi:2015cka, Bonati:2015vqz, Petreczky:2016vrs, Borsanyi:2016ksw, Burger:2018fvb, Bonati:2018blm, Bottaro:2020dqh, Lombardo:2020bvn}. A fit to the numerical results from lattice simulations is~\cite{Borsanyi:2016ksw}
\begin{equation}
    \label{eq:chi}
    \chi(T) \simeq \chi_0 \times
    \begin{cases}
        1\,, &\text{ for } T \lesssim T_\text{QCD}\,,\\
        \left(\frac{T}{T_\text{QCD}}\right)^{-b}\,, &\text{ for } T \gtrsim T_\text{QCD}\,,
    \end{cases}
\end{equation}
where $\chi_0 \simeq 0.0216\,$fm$^{-4}$ and $b \simeq 8.16$. At any temperature $T$, the mass of the axion is $m(T) = \sqrt{\chi(T)}/f_a$, so the axion can be effectively regarded as a massless scalar field as long as the QCD effects can be neglected for $T \gg T_\text{QCD}$. In the opposite limit $T \ll T_\text{QCD}$, the mass squared of the axion at zero temperature is~\cite{Weinberg:1977ma}
\begin{equation}
    m_a^2 \equiv m^2(T=0) = \frac{m_u\, m_d}{(m_u+m_d)^2}\frac{m_\pi^2\, f_\pi^2}{f_a^2}\,,
\end{equation}
where $m_u$, $m_d$ are the masses of the up and down quarks, $m_\pi \simeq 140\,$MeV is the mass of the $\pi$ meson, and $f_\pi \simeq 92\,$MeV is the pion decay constant. Numerically, this gives $m_a = \Lambda_a^2/f_a$, with $\Lambda_a = \chi_0^{1/4} \simeq 75.5\,$MeV.

The equation of motion for the axion field obtained from the Lagrangian in Eq.~\eqref{eq:lagrangian} is
\begin{equation}
    \label{eq:eom}
    \ddot \theta + 3\, H(T)\, \dot \theta - \frac{1}{R^2(T)}\nabla^2\theta + m^2(T)\, \sin\theta = 0\,,
\end{equation}
where a dot is a differentiation with respect to the cosmic time $t$, $R=R(T)$ is the scale factor, and $H(T)\equiv \dot R/R$ is the Hubble expansion rate. This expression for the evolution of the axion field is a Klein-Gordon equation in the potential of Eq.~\eqref{eq:axion_potential}. At any time, the axion energy density is
\begin{equation}
    \label{eq:energydensity}
    \rho_a(T) = \frac12\, f_a^2\, \dot\theta^2 + \frac12\, \frac{1}{R^2}\, f_a^2\,(\partial_\mu\theta)^2 + m^2(T)\, f_a^2 \left(1 - \cos\theta\right)\,.
\end{equation}

Equation~\eqref{eq:eom} is solved by evolving the axion field starting from the initial conditions $\theta = \theta_i$ and $\dot \theta = \dot \theta_i$ which are imposed when the saxion field begins to oscillate~\cite{Co:2019jts}. We refer to $\theta_i$ as the initial misalignment angle, while the initial velocity is usually set as $\dot \theta_i = 0$.\footnote{An alternative production mechanism which is valid for a large initial value of the saxion field is parametric resonance~\cite{Shtanov:1994ce, Kofman:1994rk, Kofman:1997yn, Co:2017mop}.} In the naive estimation of the axion abundance, Eq.~\eqref{eq:eom} gets solved by considering super-horizon modes for which the gradient term is negligible. More elaborate treatments have to rely on the lattice simulation of Eq.~\eqref{eq:eom} and the interaction of the axion with the radial component of the PQ field. These computations are extremely demanding and lead to conflicting results in the literature. The string network that develops in fully solving Eq.~\eqref{eq:eom} has a spectrum that spans all frequencies from the infrared cutoff $\sim H$ to the ultraviolet (UV) cutoff $\sim f_a$, with a spectral index $q$. Current simulations can explore scales down to a size in which the behavior seems to be dominated by the UV spectrum with $q < 1$~\cite{Gorghetto:2018myk}, however recent results seem to be challenged once even more refined grids are used~\cite{Gorghetto:2020qws}. In this work, we qualitatively remark the differences between the various misalignment scenarios, for which we rely on solving Eq.~\eqref{eq:eom} for super-horizon modes, and we neglect the contribution from strings.

In the misalignment mechanism, the axion field starts to roll about the minimum of the potential once the Hubble friction is overcome by the potential term~\cite{Preskill:1982cy, Stecker:1982ws, Abbott:1982af, Dine:1982ah}. This occurs around the temperature $T_\text{osc}^\text{mis}$ defined implicitly as
\begin{equation}
    \label{eq:Tosc}
    3\, H(T_\text{osc}^\text{mis}) \approx m(T_\text{osc}^\text{mis})\,,
\end{equation}
where generally $m(T_\text{osc}^\text{mis}) \ll m_a$. We assume the standard radiation-dominated phase, during which the Hubble rate is
\begin{equation}
    \label{eq:hubblerate}
    H(T) = \frac{\pi}{3} \sqrt{\frac{g_\star(T)}{10}}\, \frac{T^2}{M_P}\,,    
\end{equation}
in which $g_\star(T)$ is the number of relativistic degrees of freedom at temperature $T$~\cite{Drees:2015exa} and $M_P$ is the reduced Planck mass. With this assumption, we obtain (see, e.g. Ref.~\cite{Visinelli:2009kt})
\begin{equation}
	\label{eq:Tosc2}
	T_\text{osc}^\text{mis} \simeq 
	\begin{cases}
		\left(\sqrt{\frac{10}{\pi^2\, g_\star(T_\text{osc}^\text{mis})}}\, M_P\, m_a\right)^{1/2}\,, &  T_\text{osc}^\text{mis} \lesssim T_\text{QCD}\,, \\
		\left(\sqrt{\frac{10}{\pi^2\, g_\star(T_\text{osc}^\text{mis})}}\, M_P\, m_a\, T_\text{QCD}^{b/2}\right)^{2/(4+b)}\,, & T_\text{osc}^\text{mis} \gtrsim T_\text{QCD}\,.
	\end{cases}\nonumber
\end{equation}
For example, an axion field of mass $m_a \simeq 26\,\mu$eV would begin to oscillate at $T_\text{osc}^\text{mis} \simeq 1.23\,$GeV. In the absence of entropy dilution, the axion number density in a comoving volume after the onset of oscillations is conserved,
\begin{equation}
    \frac{\mathrm{d}}{\mathrm{d}t}\,\left[\frac{\rho_a(T)/m(T)}{s(T)}\right] = 0\,,
\end{equation}
where $s(T) = (2\pi^2/45)\, g_{\star s}(T)\, T^3$ is the entropy density and $g_{\star s}(T)$ is the number of entropy degrees of freedom at temperature $T$~\cite{Drees:2015exa}.
This last expression gives the present axion density fraction,
\begin{equation}
    \label{eq:axionfraction}
    \Omega_a = \frac{\rho_a(T_\star)}{\rho_{\rm crit}}\frac{m_a}{m(T_\star)}\,\frac{g_{\star s}(T_0)}{g_{\star s}(T_\star)}\frac{T_0^3}{T_\star^3}\,,
\end{equation}
where $T_\star$ is any temperature such that $T_\star < T_\text{osc}^\text{mis}$, $T_0$ is the present CMB temperature, and the critical density is given in terms of the Hubble constant $H_0$ as $\rho_{\rm crit} = 3\, M_P^2\, H_0^2$. The energy density in Eq.~\eqref{eq:energydensity} is approximated in the limit in which the kinetic energy can be neglected and for a quadratic potential such as
\begin{equation}
    \label{eq:rhoMis}
    \rho_a(T_\star) \simeq \frac{1}{2} m^2(T_\star)\, f_a^2\, \theta_i^2\,,
\end{equation}
where $\theta_i$ is the initial value of the misalignment angle at temperatures $T \gg T_\text{osc}^\text{mis}$. As an order of estimate, for $f_a\simeq 10^{12}\,$GeV the correct relic density that matches the observed DM is obtained for initial field values $\theta_i\simeq\mathcal{O}(1)$.\footnote{Astrophysical constraints provide a lower bound on the decay constant requiring $f_a\gtrsim 10^7\,$GeV~\cite{Raffelt:2006cw, Co:2020dya}.}

\begin{figure}
	\centering
    \includegraphics[width=0.48\textwidth]{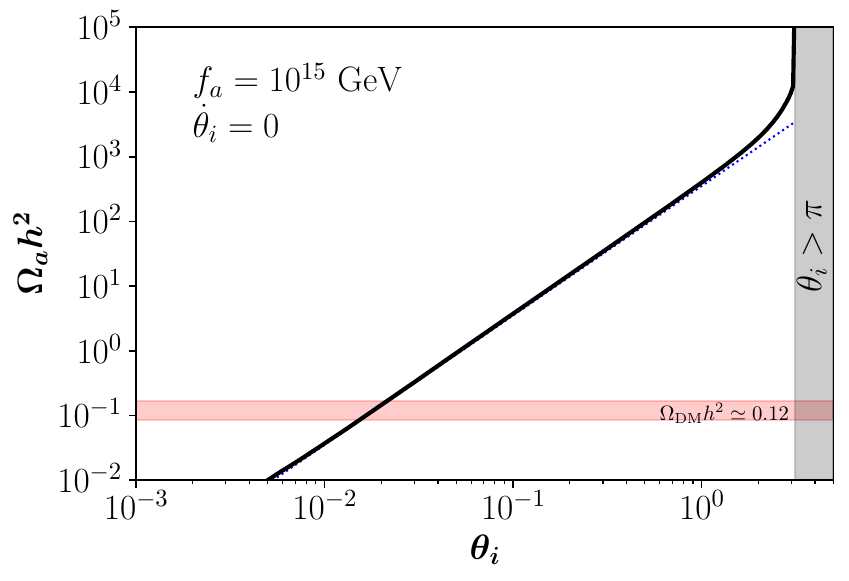}\vspace{0.25cm}
    \includegraphics[width=0.48\textwidth]{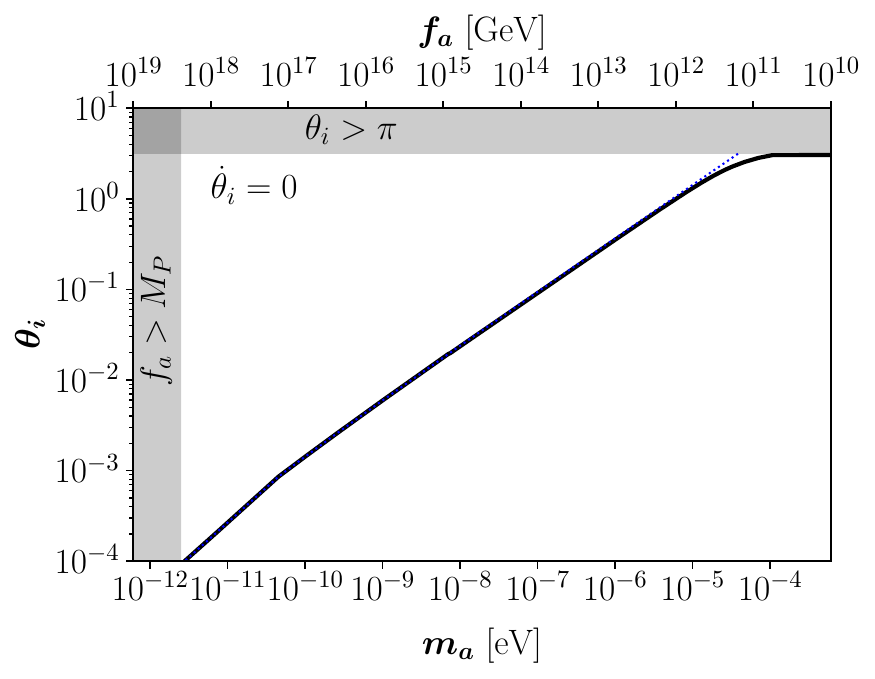}
    \caption{Standard scenario. Left panel: Axion relic abundance for $f_a = 10^{15}\,$GeV, taking $\dot\theta_i = 0$.
    Right panel: Misalignment angle required to match the whole observed DM abundance.
    The two lines show the comparison between numerical (solid black) and analytical (dotted blue).}
	\label{fig:Oh2_th}
\end{figure}

The left panel of Fig.~\ref{fig:Oh2_th} shows the axion relic abundance $\Omega_ah^2$ as a function of $\theta_i$ for $f_a = 10^{15}\,$GeV, corresponding to the axion mass $m_a \approx 5.7\,$neV, in the standard misalignment case (i.e., taking $\dot\theta_i = 0$). Since the axion potential is symmetric, hereafter without loss of generality we assume $\theta_i \geq 0$. The solid black and the dotted blue lines in Fig.~\ref{fig:Oh2_th} correspond to the results obtained from the numerical solution of Eq.~\eqref{eq:eom} and the analytical solutions in Eq.~\eqref{eq:axionfraction}, respectively. The analytical solution with the quadratic approximation in Eq.~\eqref{eq:rhoMis} underestimates the relic abundance when $\theta_i \simeq \pi$, due to the presence of the non-harmonic terms in the QCD axion potential~\cite{Lyth:1991ub, Bae:2008ue, Visinelli:2009kt, Arias:2021rer}. The red horizontal band showing to $\Omega_a h^2 \simeq \Omega_{\rm DM}h^2$, where $\Omega_{\rm DM}h^2 \approx 0.12$ is the DM abundance today from the {\it Planck} satellite measurements~\cite{Planck:2018vyg}. The right panel of Fig.~\ref{fig:Oh2_th} depicts the misalignment angle required to produce the whole observed DM abundance in the standard scenario as a function of the QCD axion mass. The slope changes for $m_a \approx 4.8\times 10^{-11}\,$eV, corresponding to the DM axion mass at which the oscillations in the axion field begins around the QCD phase transition $T_\text{osc}^\text{mis} = T_\text{QCD}$. For $f_a\gg 10^{12}\,$GeV, the initial misalignment angle must be tuned so that $f_a\,\theta_i^2$ is approximately constant to give rise to the observed abundance. For $f_a \ll 10^{12}\,$GeV, the abundance of cold axions is much smaller than that of DM unless the initial misalignment angle gets tuned to $\theta_i \approx \pi$. In this region of the parameter space, the relevant non-harmonic contributions to the QCD axion potential break the analytic derivation sketched in Eq.~\eqref{eq:rhoMis}, and a numerical solution of Eq.~\eqref{eq:eom} is needed.

\section{Kinetic misalignment mechanism}
\label{sec:misalignment}

A non-negligible initial rotations of complex scalar fields could arise through the Affleck-Dine mechanism~\cite{Affleck:1984fy}. When applied to the PQ field, baryogenesis can be addressed by transferring the PQ charge associated to the rotation of the PQ field to the SM chiral asymmetries, which in turn generate the baryon asymmetry through baryon number-violating processes such as electroweak sphaleron processes~\cite{Co:2019wyp}. For the sake of generality, we remain agnostic about the detailed mechanism that generate the initial kick of the axion and study the implication caused by such an assumption, although various scenarios have been discussed in the literature in relation with the pre-inflation QCD axion, see e.g.\ Refs.~\cite{Co:2019jts, Co:2019wyp, Chang:2019tvx, Co:2020dya, Co:2020xlh, Co:2020jtv, Harigaya:2021txz, Co:2021lkc, Co:2022qpr, Barnes:2022ren}.

%
%
Our intention with this work is neither to elaborate on how the initial kick is generated in the early Universe nor extensively work out its detailed dynamics once different patches reach causal contact. We assume that the initial kick is provided by some mechanism acting around the phase transition and we look into what observational consequences this provides.

\subsection{Overview}

So far, we have reviewed the computation of the DM abundance in the standard scenario in which the initial value of the axion field velocity gets set to zero. Recently, the possibility has been considered that the axion field possesses a non-zero initial velocity $\dot \theta_i \neq 0$, which is so large that the potential barriers can effectively be ignored, in the so-called kinetic misalignment (KM) mechanism~\cite{Co:2019jts, Chang:2019tvx}.

In the KM mechanism, the axion possesses an initial velocity $\dot{\theta}_i \neq 0$ corresponding to the rotation of the PQ field $\Phi$ in the complex field plane and an overall asymmetry of the PQ charge. At any time, the Noether charge density associated to the shift symmetry of the axion field in Eq.~\eqref{eq:shiftsymmetry} is
\begin{equation}
    \label{eq:chargedensity}
    n_{\theta} = i \left[\Phi \dot{\Phi}^{*} - \Phi^{*}\dot{\Phi}\right] = \dot{\theta}\,f_a^2\,,
\end{equation}
with the corresponding yield $Y_{\theta} \equiv \dot{\theta}f_{a}^2/s(T)$. A necessary condition to generate an initial velocity of the axion field consists in the field value of the radial mode $S$ to be initially much larger than the axion decay constant, $S \gg f_{a}$, as it occurs in the early Universe. Provided that the PQ potential is sufficiently flat, the desired conditions can be realized by either imposing the appropriate initial conditions of inflation, primordial quantum fluctuations, or in supersymmetric models with flat directions in the superpotential. Another possibility to generate the initial misalignment velocity $\dot\theta_{i} \neq 0$ can emerge from axion models where the axion potential becomes tilted by introducing an explicit symmetry breaking term induced by a higher-dimension potential of the form
\begin{equation}
    \label{eq:explicitbreak}
    V_{\rm PQ-break} = M_P^4\,\left(\frac{\Phi}{M}\right)^n + {\rm h.c.}\,,
\end{equation}
where $n > 0$ is an integer and $M$ is a new energy scale lying well beyond the SM. The addition of the potential in Eq.~\eqref{eq:explicitbreak} would lead to an explicit breaking of the PQ symmetry, and would provide an initial kick to the angular direction of $\Phi$~\cite{Co:2019jts}.

Two distinct regions exist for this mechanism: in the {\it weak} KM regime, the initial velocity allows the axion to explore a few different minima of the potential, while in the {\it strong} KM regime the initial axion velocity is so large that the potential barriers can effectively be ignored, and the onset of coherent oscillations gets delayed. In the following subsections, we discuss DM production in these different regimes.

\subsection{Weak kinetic misalignment}
\label{sec:weakkm}

At high temperatures $T \gg T_\text{osc}^\text{mis}$, the potential term in Eq.~\eqref{eq:eom} can be safely neglected and the oscillations in the axion field are damped by the Hubble friction. Therefore, the expression $\ddot \theta + 3H(T)\, \dot \theta \simeq 0$ predicts $\dot\theta \propto R^{-3}$ for any cosmological model. This result can also be obtained from the conservation of the Noether charge $n_\theta$ in Eq.~\eqref{eq:chargedensity}~\cite{Co:2019jts}. For this reason, if the field possesses a non-zero initial velocity, the kinetic energy term would dominate over the potential energy term and the total energy density would scale as a kination field~\cite{Barrow:1982ei},
\begin{equation}
    \rho_a \propto \dot\theta^2 \propto R^{-6}\,.
\end{equation}
Since in a radiation dominated cosmology $H(R) \propto R^{-2}$, the axion field in this configuration would scale as~\cite{Chang:2019tvx}
\begin{equation}
    \label{eq:int0}
    \theta(R) \simeq \theta_i - 2n\, \pi + \frac{\dot\theta(R_i)}{H(R_i)} \left[1 - \frac{R_i}{R}\right],
\end{equation}
with $n$ being a natural number that counts how many times the axion crosses a potential barrier, and $\theta_i = \theta(R_i)$ is the initial misalignment angle. Equation~\eqref{eq:int0} can be restated in terms of the the dimensionless redshift-invariant yield $Y_\theta \equiv f_a^2\, \dot\theta/s(T)$ as
\begin{equation}
    \label{eq:theta_mis}
    \theta(T) \simeq \theta_i - 2n\, \pi + \frac{Y_\theta}{f_a^2} \frac{s(T_i)}{H(T_i)} \left[1 - \left(\frac{s(T)}{s(T_i)}\right)^{1/3}\right],
\end{equation}
where $T_i$ is the photon temperature at $R = R_i$. The standard misalignment scenario in Eq.~\eqref{eq:rhoMis} is recovered in the limit $Y_\theta = 0$ and $n = 0$. In this case, $T_\text{osc}^\text{mis}$ does not vary with respect to the standard misalignment scenario and Eq.~\eqref{eq:Tosc} holds. At later times, the Hubble rate dampens the oscillations and the QCD potential becomes relevant.

In the weak KM regime, the value of the misalignment angle $\theta_i$ in Eq.~\eqref{eq:int0} required to match the observed DM abundance is modified by the presence of the initial velocity term $\dot\theta(R_i)/H(R_i)$. The parameter space for the KM mechanism is non-trivial and depends on the direction of the velocity $\dot{\theta}_i$ which changes the sign of the yield $Y_\theta$.

\begin{figure}[t!]
	\centering
    \includegraphics[width=0.48\textwidth]{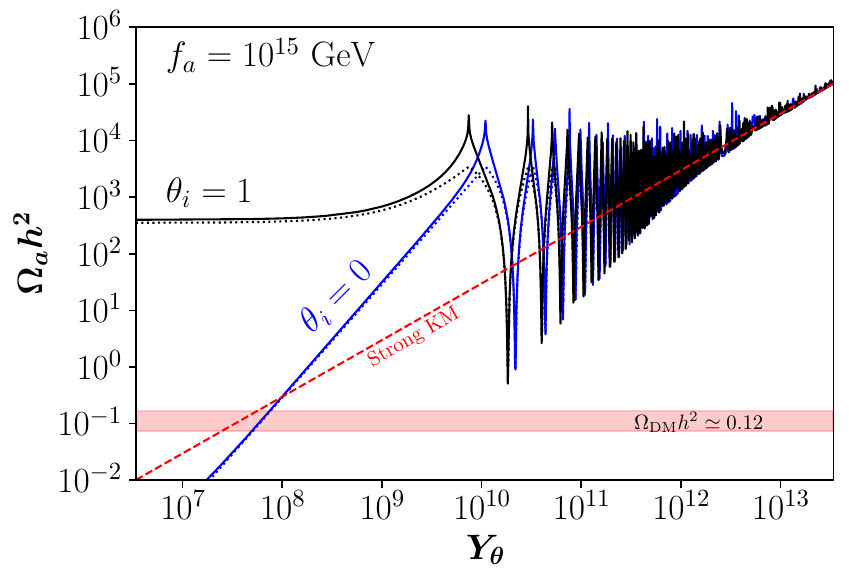}\vspace{0.25cm}
    \includegraphics[width=0.48\textwidth]{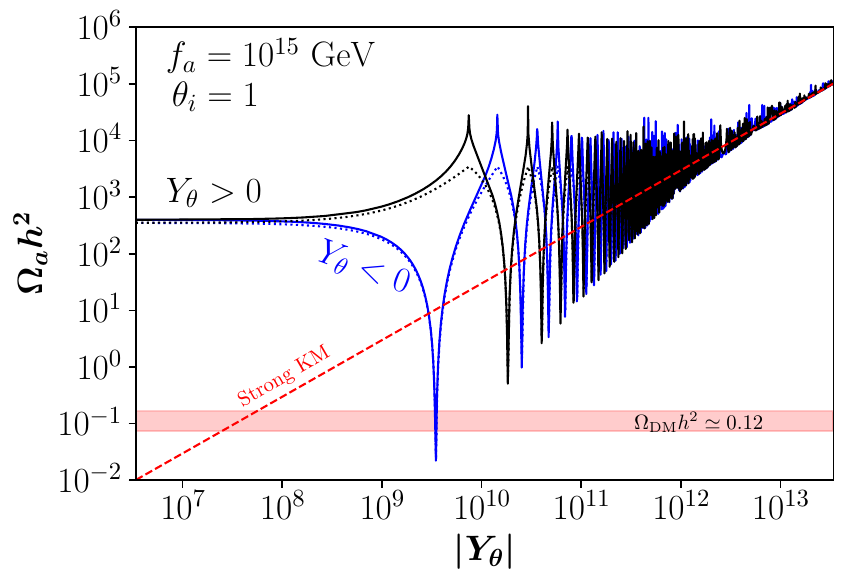}
    \caption{Weak KM. Axion relic abundance for $f_a = 10^{15}\,$GeV.
    Left panel: $\theta_i = 1$ (black) and $\theta_i = 0$ (blue) for $Y_\theta > 0$. Right panel: $Y_\theta >0$ (black) and $Y_\theta <0$ (blue), for $\theta_i = 1$. We show a comparison between numerical results (solid) and the analytical approximation with a quadratic QCD axion potential (dotted). The dashed red line shows the limit of strong KM.}
	\label{fig:Oh2_thp}
\end{figure} 
Fig.~\ref{fig:Oh2_thp} (left panel) shows the dependence of the axion relic abundance $\Omega_ah^2$ on the redshift-invariant yield $Y_\theta > 0$, for a fixed PQ breaking scale $f_a = 10^{15}\,$GeV and for the choices $\theta_i = 1$ (black) or $\theta_i = 0$ (blue). The solid and the dotted lines correspond to the numerical solution and its analytical approximation obtained by considering a quadratic QCD axion potential instead of Eq.~\eqref{eq:axion_potential}. The approximation is in good agreement with the numerical result unless for values of the misalignment angle in Eq.~\eqref{eq:theta_mis} near $\theta \simeq (1 + 2 n) \pi$, corresponding to the peaks in the cosine potential. For small initial velocities, the axion energy density is dominated by the potential, therefore a larger initial misalignment angle $\theta_i \approx 1$ gives rise to a higher relic abundance as in the case of standard misalignment. However, when the kinetic energy starts to dominate, the energy density rapidly grows as $\Omega_a h^2 \propto Y_\theta^2$. This behavior halts once the initial kinetic energy is large enough so that the field climbs the top of the potential, $\theta = (1 + 2 n) \pi$. A higher value for $Y_\theta$ allows the crossing of the potential barrier, oscillations taking place when the axion starts to roll down the potential, and hence a smaller relic abundance in generated. A minimum for $\Omega_a h^2$ occurs when oscillations start at the minimum of the potential where $\theta = 2n\, \pi$. Increasing values for $Y_\theta$ allow the axion field excursion to cross several potential barriers, and therefore the relic abundance experiences the oscillating behavior shown in Fig.~\ref{fig:Oh2_thp}. Furthermore, when the kinetic energy completely dominates over the potential, $\Omega_a h^2$ loses its dependence on the initial misalignment angle $\theta_i$, and it asymptotically approaches the strong KM regime, see Eq.~\eqref{eq:rhoKM} in the following section. The right panel of Fig.~\ref{fig:Oh2_thp} shows the results for $\Omega_ah^2$ once fixing $\theta_i = 1$ and for the yields $Y_\theta >0$ (black) or $Y_\theta <0$ (blue). For positive values of the misalignment angle, the choices $Y_\theta > 0$ and $Y_\theta < 0$ correspond to an axion climbing up or rolling down further in the potential well, respectively. We emphasize that in the case of weak KM, the oscillation temperature is the same as in the standard misalignment scenario, so that there is no delay the onset of oscillations.

\begin{figure}
	\centering
    \includegraphics[width=0.48\textwidth]{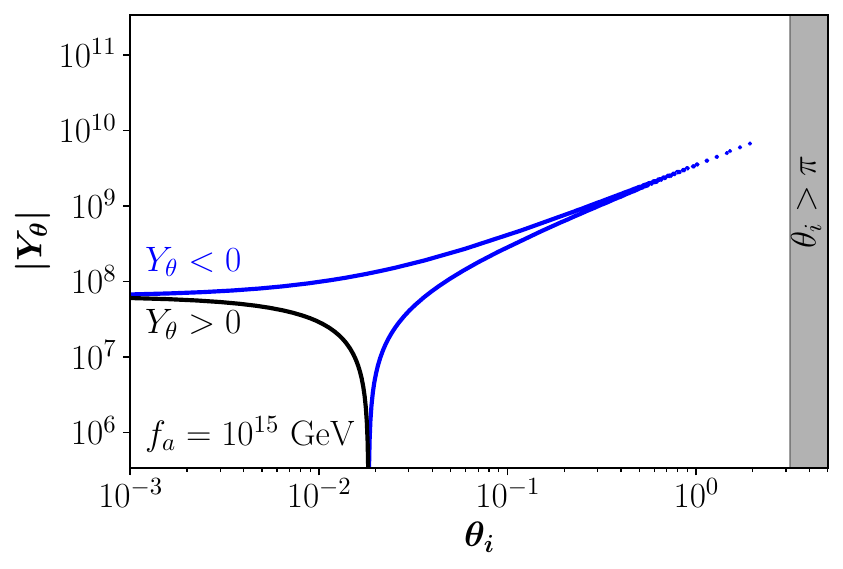}\vspace{0.25cm}
    \includegraphics[width=0.48\textwidth]{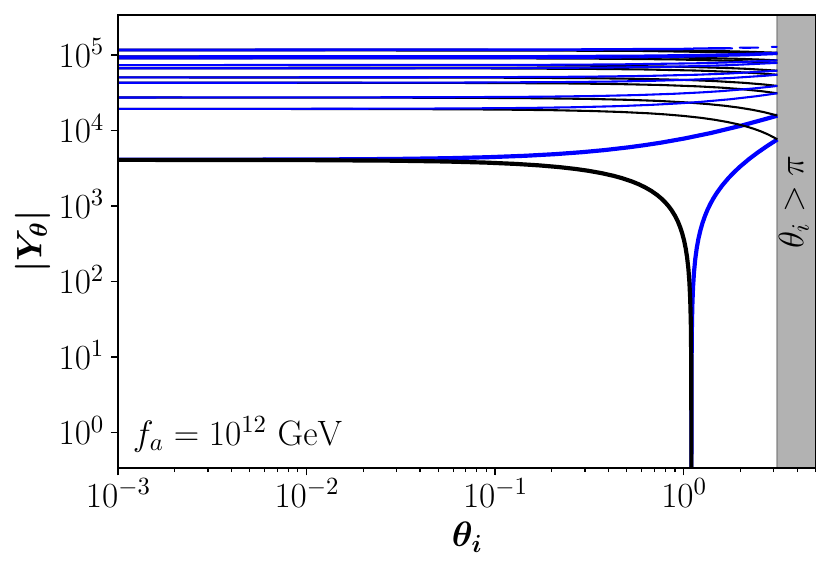}\\
    \caption{Kinetic misalignment. Parameter space that reproduces the whole observed DM abundance for $f_a = 10^{15}\,$GeV (Left) and $f_a = 10^{12}\,$GeV (Right).
    Black and blue lines correspond to $Y_\theta > 0$ and $Y_\theta < 0$, respectively.
    }
	\label{fig:kinmis}
\end{figure}

We now fix the relic abundance of axions to that of the observed DM, and study the corresponding parameter space $\{ \theta_i, \lvert Y_\theta\rvert\} $ for which this is achieved. This is shown in Fig.~\ref{fig:kinmis} for the different choices $f_a = 10^{15}\,$GeV (left panel) and $f_a = 10^{12}\,$GeV (right panel). In both panels, the black lines correspond to $Y_\theta > 0$, while blue lines correspond to $Y_\theta < 0$. For $f_a = 10^{15}\,$GeV, the axion is confined in the potential well containing its minimum and it is not able to explore other minima, i.e.\ there are only solutions corresponding to $n=0$ in Eq.~\eqref{eq:int0}. For $Y_\theta < 0$, the solution features a spike-like behavior, corresponding to the first funnel-shaped region appearing in the right panel of Fig.~\ref{fig:Oh2_thp}. In the case $Y_\theta<0$, the axion field has a negative moderate initial velocity that makes it roll down further in the potential well so that the field value becomes smaller than $\theta_i$ when the oscillations begin; this leads to a suppression in the relic abundance and, as a consequence, a larger $\theta_i$ is required to compensate.
A similar behavior is shown for $f_a = 10^{12}\,$GeV, however in this case different solutions to Eq.~\eqref{eq:int0}, corresponding to higher values of $n$, give rise to the observed DM abundance. In this scenario, the axion has enough kinetic energy to explore different minima, and therefore different solutions corresponding to the same initial axion angle appear. As the total axion energy density is dominated by the kinetic term, the new solutions tend to be independent of $\theta_i$.

\begin{figure}[ht]
	\centering
    \includegraphics[width=0.48\textwidth]{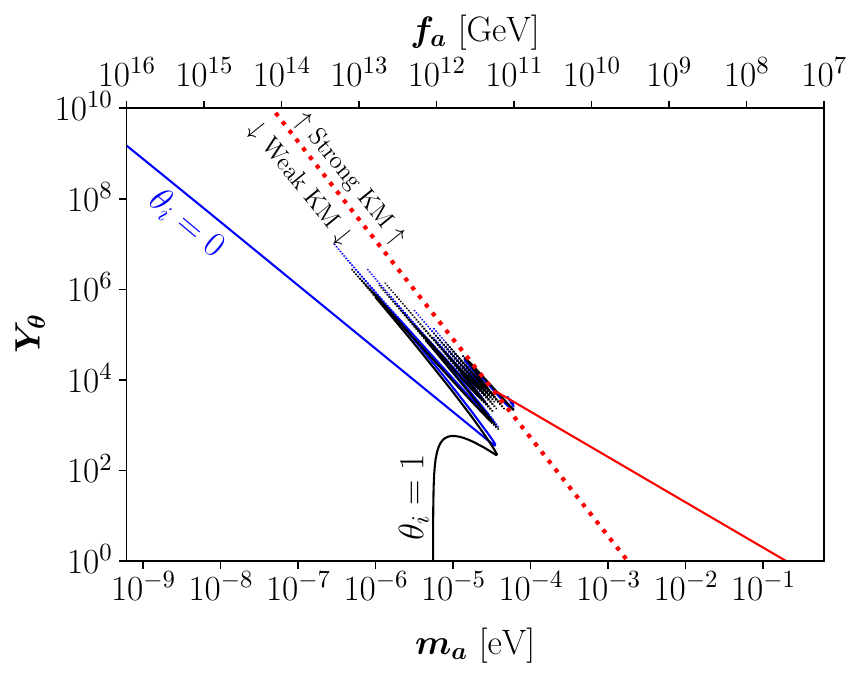}\vspace{0.25cm}
    \includegraphics[width=0.48\textwidth]{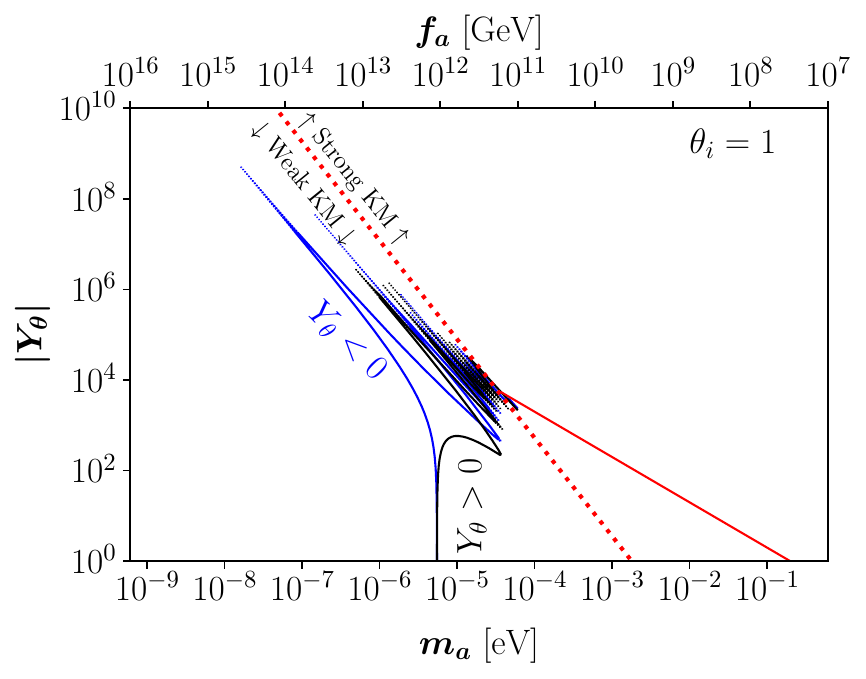}
    \caption{Weak kinetic misalignment. Left panel: Yield $Y_\theta$ as a function of the DM axion mass $m_a$, for the values of the initial misalignment angle $\theta_i = 0$ (blue) and $\theta_i = 1$ (black). Right panel: $\lvert Y_\theta\rvert$ as a function of the DM axion mass $m_a$,   assuming $\theta_i = 1$ and considering either $Y_\theta > 0$ (black) or $Y_\theta < 0$ (blue). In both panels, the red dotted line separates the regimes of weak (to left) and strong (to right) KM.}
	\label{fig:Y-ma}
\end{figure}

A similar behavior occurs when plotting the contours describing axion DM over the plane $\{m_a,\, Y_\theta\}$, see Fig.~\ref{fig:Y-ma}. The left panel shows the region $Y_\theta > 0$ for the initial axion angles $\theta_i = 0$ (blue) and $\theta_i = 1$ (black), whereas the right panel shows the cases for $Y_\theta > 0$ (black) and $Y_\theta < 0$ (blue) assuming $\theta_i = 1$. In order to reproduce the same observed abundance in the KM scenario, the case for $\theta_i = 0$ requires a larger value of $Y_\theta$ compare to the case $\theta_i = 1$. The dotted lines marks the area, to the left of the line, for which the weak KM regime holds, while the strong KM regime applies to the right, see Eq.~\eqref{eq:skm} below. All solutions with different initial values of $\theta_i$ (left panel) or the yield $Y_\theta$ (right panel) converge to the solution given by the red solid line in the strong KM regime.

\subsection{Strong kinetic misalignment}
\label{sec:strongkm}

Contrary to the previous case where an initial velocity allows the axion to explore a few minima, in the strong KM the kinetic energy is so large compared to the potential barrier that the potential is effectively flat. In this scenario, the oscillations in the axion field are delayed with respect to the cases of the standard misalignment and weak KM~\cite{Co:2019jts, Chang:2019tvx}. The initially dominant axion kinetic energy $K = \dot a^2/2$ eventually becomes equal to the maximum of the potential barrier $V_\text{max} = 2\, m^2(T)\, f_a^2$ at the temperature $T_\text{osc}^\text{skm}$, defined implicitly by the equality
\begin{equation}
	\label{eq:Tk}
	\lvert \dot\theta(T_\text{osc}^\text{skm})\rvert \equiv 2\, m(T_\text{osc}^\text{skm})\,.
\end{equation}
If at $T = T_\text{osc}^\text{mis}$ the kinetic energy density is larger than the potential barrier, the axion oscillations are delayed until the kinetic energy falls below the potential energy. With $m(T_\text{osc}^\text{mis}) \approx 3H(T_\text{osc}^\text{mis})$, this condition is satisfied for
\begin{equation}
    \lvert \dot\theta(T_\text{osc}^\text{mis})\rvert = \lvert\dot\theta(T_i)\rvert\, \frac{s(T_\text{osc}^\text{mis})}{s(T_i)} = \frac{\lvert Y_\theta\rvert}{f_a^2}\, s(T_\text{osc}^\text{mis}) \gtrsim 6\, H(T_\text{osc}^\text{mis})\,,
\end{equation}
which corresponds to the red dotted line appearing in the panels of Fig.~\ref{fig:Y-ma}. In terms of the yield, this gives
\begin{equation}
    \label{eq:skm}
    \lvert Y_\theta\rvert > 6\, f_a^2\, \frac{H(T_\text{osc}^\text{mis})}{s(T_\text{osc}^\text{mis})}\,.
\end{equation}

To obtain the present relic abundance, we employ the conservation of $n(T) / s(T)$ from the onset of field oscillations to present time,
\begin{align}
    \label{eq:rhoKM}
    \rho_a(T_0) &= \mathcal{C}\, \rho_a(T_\text{osc}^\text{skm}) \frac{m_a}{m(T_\text{osc}^\text{skm})} \frac{s(T_0)}{s(T_\text{osc}^\text{skm})}\nonumber\\
    &\simeq \mathcal{C}\, \lvert \dot\theta(T_\text{osc}^\text{skm})\rvert\, f_a^2\, m_a\, \frac{s(T_0)}{s(T_\text{osc}^\text{skm})}
    = \mathcal{C}\, \lvert Y_\theta\rvert\, m_a\, s(T_0)\,,
\end{align}
where we used the fact that in the strong KM scenario the axion energy density is completely dominated by the kinetic energy. Although the analytical estimate predicts $\mathcal{C} = 1$, a numerical analysis favors $\mathcal{C} \simeq 2$~\cite{Co:2019jts}. The result in Eq.~\eqref{eq:rhoKM} is the red dashed line in Fig.~\ref{fig:Oh2_thp}, and the red solid lines in Fig.~\ref{fig:Y-ma}. As evident from Eq.~\eqref{eq:rhoKM}, in this limit the axion DM relic abundance is independent of the initial misalignment angle and the sign of $Y_\theta$. The transition between the weak and the strong KM regimes occurs at $f_a \simeq 2.2 \times 10^{11}\,$GeV, corresponding to $m_a \simeq 26\,\mu$eV.

\begin{figure}[t!]
	\centering
	\includegraphics[width=0.8\textwidth]{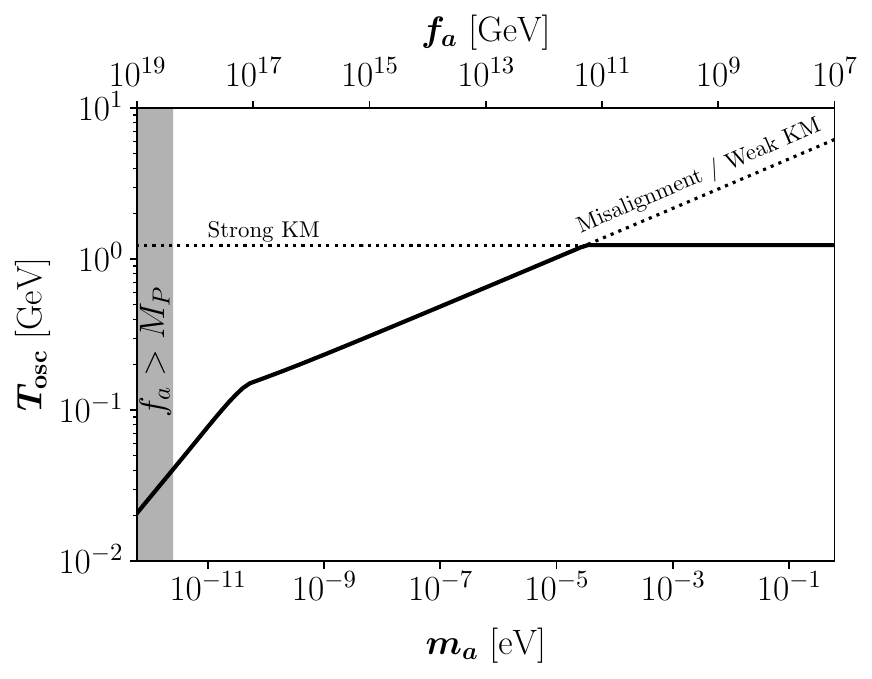}
	\caption{The temperature of the primordial plasma at the onset of axion field oscillations $T_{\rm osc}$ as a function of the axion mass. The tilted dotted line corresponds to the cases of standard misalignment or weak KM regimes in Eq.~\eqref{eq:Tosc2}, while the horizontal dotted line is the result in the strong KM regime in Eq.~\eqref{eq:Toscskm}.}
	\label{fig:Tosc}
\end{figure}

The KM mechanism allows us to explore the region of the parameter space corresponding to relatively large values of the axion mass.\footnote{Axions heavier than $\mathcal{O}(10^{-1})\,$eV are in tension with observations from horizontal branch stars and other astrophysical measurements~\cite{Ayala:2014pea}.} Equations~\eqref{eq:Tk}--\eqref{eq:rhoKM} allow us to compute the value of $T_\text{osc}^\text{skm}$ required to match the observed DM abundance, via the relation
\begin{equation}
    \label{eq:Tskm}
    \frac{\sqrt{\chi(T_\text{osc}^\text{skm})}}{s(T_\text{osc}^\text{skm})} \simeq \frac{\rho_{\rm DM}}{2\, \mathcal{C}\, s(T_0)\, \sqrt{\chi(T_0)}}\,,
\end{equation}
which, setting $\mathcal{C}=2$, yields a value that is independent on the axion mass,
\begin{equation}
    \label{eq:Toscskm}
    T_\text{osc}^\text{skm} \!=\! \left[\frac{4\,g_{\star s}(T_0)}{g_{\star s}(T_\text{osc}^\text{skm})} \frac{\chi_0}{\rho_{\rm DM}}\, T_{\rm QCD}^{b/2}\, T_0^3\right]^{\frac{2}{6 + b}} \!\simeq\! 1.23{\rm \,GeV}\,.
\end{equation}
Thus, in the strong KM scenario, axions start to oscillate at a smaller temperature $T=T_\text{osc}^\text{skm}$ instead of the value obtained in the conventional scenario $T_\text{osc}^\text{mis}$ given in Eq.~\eqref{eq:Tosc2} and, as a consequence, the onset of coherent oscillations is delayed.\footnote{A delay in the onset of oscillations also occurs in the {\it trapped misalignment} case~\cite{Nakagawa:2020zjr, DiLuzio:2021gos}.}
We have shown the value of the temperature at which the axion field is set into motion as a function of its mass in Fig.~\ref{fig:Tosc}. The line denoted ``Strong KM'' is the value given in Eq.~\eqref{eq:Toscskm}, and the two tilted lines to the left denote the result in Eq.~\eqref{eq:Tosc2}. A summary of the conditions satisfied by the three different misalignment mechanisms discussed is given in Table~\ref{tab:mis-cond}.
\begin{table}[htb!]
\def\arraystretch{1.5}
\centering
\begin{tabular}{|l||c|c|}
\hline
Mechanism & Initial velocity & Oscillation temperature\\
\hline \hline
Standard scenario & $\dot{\theta}_i=0$ & $3H(T_\text{osc}^\text{mis}) = m(T_\text{osc}^\text{mis})$ \\ \hline
Weak KM & $\dot{\theta}_i\neq$ 0 & $3H(T_\text{osc}^\text{mis}) = m(T_\text{osc}^\text{mis})$ \\ \hline
\multirow{2}{*}{Strong KM} & \multirow{2}{*}{$\dot{\theta}_i \neq 0$} & $\lvert\dot\theta(T_\text{osc}^\text{skm})\rvert = 2\, m(T_\text{osc}^\text{skm})$ \\
&  & with $\lvert \dot\theta(T_\text{osc}^\text{mis})\rvert > 2\, m(T_\text{osc}^\text{mis})$\\ 
\hline
\end{tabular}
\caption{Conditions for the various scenarios of misalignment mechanism discussed in the text.}
\label{tab:mis-cond}
\end{table}

\section{Axion miniclusters}
\label{sec:minicluster}

An axion minicluster is a dense, virialized clump of axions described by a mass $M_0$ and an overdensity $\delta$ parametrizing the local overdensity in the axion energy density. AMCs are generally discussed within the post-inflation framework, in which the PQ symmetry is spontaneously broken after inflation and the axion field at the time of the onset of oscillations is spatially inhomogeneous over different Hubble patches. However, such a scenario is difficult to realize as the KM is easily accommodated within the inflation period and due to the properties of the PQ potential in Eq.~\eqref{eq:mexicanhat}. In more detail, if the PQ scalar field is driven to large values after the post inflationary PQ breaking, its angle will be randomized over different Hubble patches, leading to a net $Y_\theta \approx 0$ and to an absence of KM effects in this picture. In addition, since the initial velocity $\dot\theta_i$ cannot exceed the mass of the radial mode $m_S = \lambda_\Phi f_a$, where the coupling is bound to be $\lambda_\Phi^2 \leq 4\pi$ by perturbativity arguments, the energy scale $f_a$ is severely constrained in this scenario. At the same time, the argument does not hold if the PQ symmetry is spontaneously broken before or during inflation and it is not restored afterwards (the pre-inflation scenario), because the assumption that $H \propto R^{-2}$ does not apply.

For the reasons above, we focus on the pre-inflation scenario, in which the axion field is homogeneous over the observable Universe and would not generally exhibit defects or seeding substructures. In this scenario, the axion field experiences the same initial conditions across the whole observable Universe; in particular, the axion would have the same initial velocity potentially leading to KM. Nevertheless, miniclusters can arise if additional features are considered. For example, a subdominant population of primordial black holes could trigger the nucleation of axion overdensities around them~\cite{Hertzberg:2020hsz, Schiappacasse:2021zlr}. Overdensities with $\delta \sim \mathcal{O}(1)$ can also arise from tachyonic instability and/or resonance instability if the axion potential contains a small explicit breaking term~\cite{Fukunaga:2020mvq, Kitajima:2021inh}, although the present analyses have been carried out only for the standard scenario and in the linear regime.

In the scenario depicted above, a sizable overdensity can be formed when the axion field starts rolling close to the hilltop of the potential~\cite{Arvanitaki:2019rax}. This is a configuration that is accompanied with the formation of domain walls not attached to strings and thus extremely dangerous cosmologically, even for $N_{\rm DW} = 1$~\cite{Linde:1990yj, Lyth:1991ub}. Nevertheless, the fine tuning required for the initial conditions is relaxed once the thermal effects from the interactions with the QCD sector are taken into account~\cite{Kitajima:2021inh}, so that axion clumps also form for a wider range of initial conditions avoiding the fine tuning. This is consistent with the requirements imposed on the axion decay constant $f_a$: in this regime, quantum fluctuations induce perturbations in the axion field of size $\sigma_a = H_I/(2\pi)$, where $H_I$ is the Hubble scale during inflation, so that the PQ symmetry is restored whenever $\sigma_a > f_a$~\cite{Linde:1990yj, Lyth:1991ub}. However, this region is not within the values of interest in this analysis, since we consider a relatively large axion decay constant $f_a \sim 10^{15}$ GeV while the energy scale of inflation for a single-field slow roll model is bound as $H_I \lesssim 2.5 \times 10^{-5}~M_P$ at 95\% confidence level (CL)~\cite{Planck:2018jri}.


\subsection{Formation and properties}

The energy density associated with an AMC at formation is~\cite{Kolb:1994fi}
\begin{equation}
    \label{eq:rhomc}
    \rho_{\rm AMC} \approx 140\, (1+\delta)\, \delta^3\, \rho_{\rm eq}\,,
\end{equation}
where $\rho_{\rm eq}$ is the energy density in DM at MRE. The comoving size of the fluctuations at the onset of oscillations is $r_H = 1/(R\, H)_{\rm osc}$,\footnote{In the following, we generally refer to $T_\text{osc}$ to indicate either Eq.~\eqref{eq:Tosc} or Eq.~\eqref{eq:Tk}.} leading to an AMC of radius $r_{\rm eq} = r_H\, R_{\rm eq}/\delta$ at the time when the overdensity perturbations decouple from the Hubble expansion and start growing by gravitational instability, rapidly forming gravitationally bound objects~\cite{Hogan:1988mp, Kolb:1993zz, Kolb:1993hw, Kolb:1994fi, Visinelli:2018wza, DiLuzio:2020wdo}, with the corresponding mass
\begin{equation}
    \label{eq:MCmass}
    M_0 = \frac{4\pi}{3} (1+\delta)\, \rho_{\rm DM}\, \frac{s(T_\text{osc})}{s(T_0)} \left(\frac{1}{H(T_\text{osc})}\right)^3.
\end{equation}
The mass scale in Eq.~\eqref{eq:MCmass} corresponds to the heaviest AMCs that are formed at MRE. Bound structures are formed with all masses below $M_0$, down to the smallest physical scales at which the oscillatory behavior of the axion field exerts an effective ``quantum'' pressure which prevents further clumping. This so-called Jeans length $\lambda_J = 2\pi/\lambda_J$ corresponds to the wave number~\cite{Khlopov:1985jw, Hu:2000ke}
\begin{equation}
    k_J = \left(16\pi G \rho_{\rm DM} m_a^2 R\right)^{1/4} \simeq 710\left(\frac{m_a}{\mu\text{eV}}\right)^{1/2}{\rm pc^{-1}},
\end{equation}
where the last expression holds at MRE. Perturbations grow for modes $k > k_J$.

Numerically, Eq.~\eqref{eq:MCmass} in different regimes reads
\begin{equation}
    \label{eq:AMCmass}
    M_0 =
    \begin{cases}
    	1.7\times 10^{-14}\,M_\odot\, (1 + \delta)\,\left(\frac{m_a}{\mu\text{eV}}\right)^{-3/2}, & \text{ for }  T_{\rm osc} \lesssim T_\text{QCD}\,,\\
    	1.1\times 10^{-10}\,M_\odot\, (1 + \delta)\, \left(\frac{m_a}{\mu\text{eV}}\right)^{-\frac{6}{4 + b}}, & \text{ for }  T_\text{QCD} \lesssim T_{\rm osc} \lesssim T_\text{osc}^\text{skm}\,,\\
    	2.1\times 10^{-11}\,M_\odot\, (1 + \delta)\,, & \text{ for }  T_{\rm osc} \gtrsim T_\text{osc}^\text{skm}\,,
    \end{cases}
\end{equation}
where the first two lines are found in the case where the onset of oscillations is not delayed by the presence of KM so that Eq.~\eqref{eq:Tosc2} holds, while the third line is obtained in the strong KM regime where Eq.~\eqref{eq:Toscskm} holds.

The first two lines of Eq.~\eqref{eq:AMCmass} apply in the pre-inflation scenario which, although less explored, could nevertheless lead to the formation of miniclusters as discussed in Sec.~\ref{sec:misalignment}. In fact, the first line in Eq.~\eqref{eq:AMCmass} corresponding to the case $T_{\rm osc} < T_\text{QCD}$ is realized for an initial misalignment angle $\theta_i \ll 1$ which, in the pre-inflation scenario, corresponds to relatively large values of $f_a$. This is consistent with the requirement that the Hubble rate at the end of the inflation epoch for single-field inflation models is bound as $H_I \lesssim 2.5 \times 10^{-5}\,M_P$ at 95\% CL~\cite{Planck:2018jri} at the wave number $k_0 = 0.002{\rm \, Mpc^{-1}}$ which, together with the bound $f_a \gtrsim H_I/(2\pi)$ valid for the pre-inflationary scenario, implies $m_a \lesssim 0.6\,\mu$eV.
The KM regime allows achieving different values of the DM axion mass according to the initial velocity, with a wider possibility for the AMC mass ranges. In particular, AMCs can be as heavy as $\sim 10^{-9}\,M_\odot$. This effect is ultimately due to the delayed onset of oscillations occurring in the KM regime, where the AMC mass is independent of the DM axion mass since the temperature in Eq.~\eqref{eq:Toscskm} is constant.

\begin{figure}[t]
    \def\sepf{0.7}
	\centering
	\includegraphics[width=0.8\textwidth]{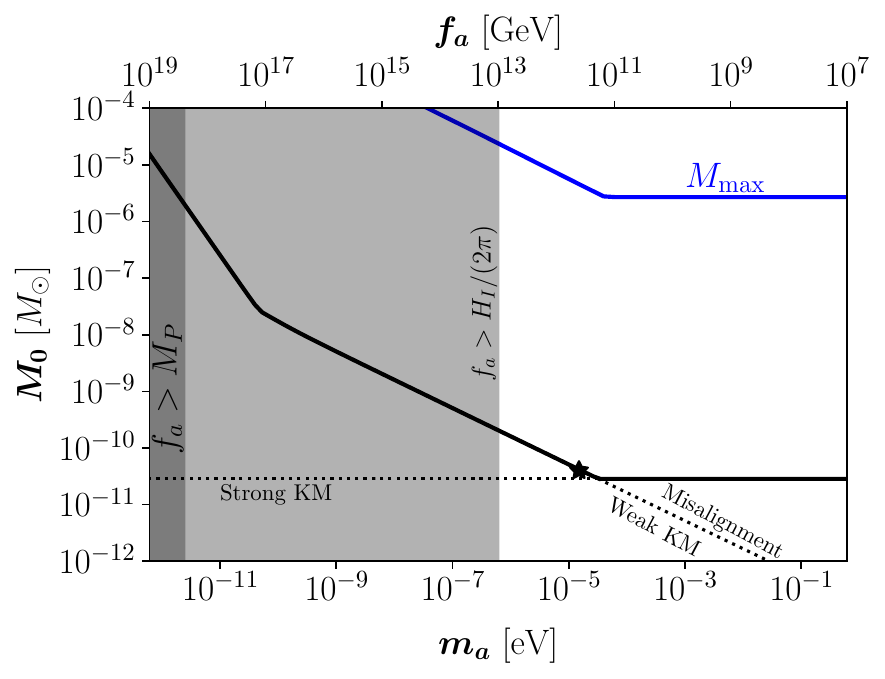}
	\caption{The characteristic minicluster mass, for $\delta = 1$. The two dotted lines correspond to the cases of standard scenario or the weak KM regime, and to the strong KM regime.}
	\label{fig:Minicluster}
\end{figure} 
The results in Eq.~\eqref{eq:AMCmass} for the AMC mass $M_0$ are sketched as a function of the DM axion mass $m_a$ by the black solid in Fig.~\ref{fig:Minicluster}. The star labels the typical AMC mass obtained with a DM axion of mass $m_a = 20\,\mu$eV using the second line in Eq.~\eqref{eq:AMCmass}. The first kink to the left corresponds to the change in the behavior of the susceptibility in Eq.~\eqref{eq:chi} near $T\sim T_\text{QCD}$, while the second kink at heavier axion masses corresponds to the change from the weak to the strong KM regimes, see Fig.~\ref{fig:Tosc}. The moment where the axion field begins to oscillate coincides with the transition from being frozen at the configuration $\theta = \theta_i$ to an oscillatory behavior with dust-like equation of state $w_a \approx 0$.

In the KM scenario, the axion initially behaves as a kination field with $w_a \approx 1$ and the transition could occur at lower temperatures, i.e.\ at $T = T_\text{osc}^\text{skm} < T_\text{osc}^\text{mis}$, as discussed in Sec.~\ref{sec:strongkm}. For this reason, axion miniclusters formed in the strong kinetic regime are heavier than those formed in the standard scenario. The gray band to the left of the plot marks the mass region in the post-inflationary scenario which is excluded by the non-observation of tensor modes in single-field inflation by the {\it Planck} satellite collaboration.

The lighter gray band marks the region where the bound $f_a > H_I/(2\pi)$ applies, which is a condition that describes the breaking of the PQ symmetry occurring during inflation. Note that, this bound applies for the quartic coupling $\lambda_\Phi = \mathcal{O}(1)$ appearing in the ``Mexican hat'' potential in Eq.~\eqref{eq:mexicanhat} that is invoked to parametrize the symmetry breaking of the PQ field. Since $\lambda_\Phi = m_S^2/f_a^2$, where $m_S$ is the mass of the radial mode $S$, this requirement applies when $m_S$ is of the same order as $f_a$. For lighter modes, the PQ symmetric phase can alternatively be obtained by adding a term describing a non-minimal interaction $\xi\, R\, |\Phi|^2$ to the Lagrangian in Eq.~\eqref{eq:lagrangian}, where $\xi$ is the non-minimal coupling strength, along with a new scale $M$ that couples to the Ricci scalar so that $M^2 = M_{\rm Pl}^2 - \xi\, |\Phi|^2$ (see Refs.~\cite{Folkerts:2013tua, Ballesteros:2016xej, Hamaguchi:2021mmt} for applications of the non-minimal coupling).

So far, the treatment has overlooked the role of density fluctuation growth, which is referred in the literature as the ``fragmentation'' of the field. It has been shown that fragmentation might play an important role in the dynamics, as it introduces an additional scale which could alter the description above~\cite{Eroncel:2022vjg}. For a model of the axion-like particle, fragmentation occurs in a sizable part of the parameter space, leading to heavier miniclusters than in the standard scenario~\cite{Eroncel:2022efc}. Fragmentation is expected to occur also for the QCD axion~\cite{Eroncel:2022vjg}, although a study of this is not yet available. The picture is complicated by the fact that the analytical tools used to study the linear regime could not suffice when the axion mass largely exceeds the Hubble rate at the trapping temperature in Eq.~\eqref{eq:Tk} (for which fragmentation is ``complete''), and a more sophisticated analysis in terms of lattice computation has to be invoked.

\subsection{Growth of structures}

At around MRE, axion miniclusters form at scales below the threshold in Eq.~\eqref{eq:MCmass}, populating the decades in mass according to a halo mass function (HMF) ${\rm d}n/{\rm (d}\ln M)$ which provides the number density $n$ as a function of the logarithmic mass $M$. The bottom-up clumping of axion miniclusters begins already around the time of matter-radiation equality when these structures form. Recent progress on the merging process has focused on the formation of axion ``minihalos'' with a HMF derived from N-body simulations~\cite{Eggemeier:2019khm, Xiao:2021nkb}, which can be understood in terms of semi-analytical modeling~\cite{Zurek:2006sy, Fairbairn:2017sil, Enander:2017ogx, Ellis:2020gtq}, and is well approximated by using the standard Press-Schechter (PS) and Sheth-Tormen formalisms.

Even though numerical simulations are in place to be able to make confident predictions, the intention here is to draw attention to a scenario that could serve as a motivation for N-body simulations in the future. Here, we intend to estimate how the HMF becomes modified in the KM case to the standard case, following the PS formalism~\cite{Press:1973iz}. The PS formalism is based on two key assumptions: $i)$ at any time, the density contrast of a spherically-symmetric overdense region of size $R$ collapses into a virialized object once it evolves above a critical overdensity $\delta_c$. For the critical value of the linear density contrast for spherical collapse during the matter domination is $\delta_c \approx 1.686$. To quantify this criterion, we introduce the overdensity fuzzed over the spherical region,
\begin{equation}
    \delta_s({\bf x},t) = \int {\rm d}^3 {\bf x'}\, \delta({\bf x'})\, W({\bf x}+{\bf x'}, R)\,,
\end{equation}
where $W({\bf x},R)$ is a kernel function that smooths the spatial overdensity over the spherical region of radius $R$. $ii)$ The density contrast is distributed as a normal distribution, specified by the variance
\begin{equation}
	\label{eq:varianceR}
	\sigma^2(z, R) = \int \frac{\mathrm{d}^3{\bf k}}{(2\pi)^3}\, \lvert \delta_k(R)\rvert^2\, \mathcal{T}^2({\bf k},z)\, \lvert W({\bf k},R)\rvert^2\,,
\end{equation}
where $\lvert \delta_k(R)\rvert^2$ is the power spectrum of the fluctuations, $\mathcal{T}({\bf k},z)$ is the transfer function, and $W({\bf k},R)$ is the Fourier transform of the kernel function $W({\bf x},R)$ that smooths the density field $\delta({\bf x})$ over the spherical region of radius $R$.

The HMF derived from these premises is parametrized as
\begin{equation}
    \frac{{\rm d}n}{{\rm d}\ln M} = \frac{\rho_{\rm DM}}{M}\,f(\sigma)\, \lvert\frac{{\rm d} \sigma}{{\rm d} \ln{M}}\rvert\,,
\end{equation}
where the multiplicity function $f(\sigma)$ is defined within the PS formalism as \begin{equation}
    f(\sigma) = \sqrt{\frac{2}{\pi}}\, \frac{\delta_{c}}{\sigma}\, e^{-\frac12 \frac{\delta_{c}^{2}}{\sigma^{2}}}.
\end{equation}
The method described above is extremely efficient in describing the distribution of the high-end mass spectrum of the HMF. Since our interest is in the heaviest virialized objects that form within the theory, we specialize the treatment to find an approximate solution for the maximal mass of the minihalo $M_{\rm max}(z)$ at redshift $z$. The power spectrum of the fluctuations $\lvert\delta_k(R)\rvert^2$ is accessible from lattice simulations in the early Universe~\cite{Vaquero:2018tib, Buschmann:2019icd} and from N-body simulations~\cite{Eggemeier:2019khm}. Since we are interested in estimating the maximal mass of axion structures at redshift $z$, we adopt the approximation in Ref.~\cite{Fairbairn:2017sil} of a white-noise power spectrum truncated at the comoving scale $k_{\rm osc}$ at which coherent field oscillations begin. Normalization of the power spectrum ensures that the integral of the power spectrum equates the square of spatial fluctuations averaged over different horizons~\cite{Peacock:1999ye} and it is here set as $P_0 = (24\pi^2/5)\, k_{\rm osc}^{-3}$~\cite{Fairbairn:2017sil}. In principle, the transfer function depends on the relative value of $k$ for the Jeans wavelengths at MRE and today. In practice, these Jeans lengths are too small to yield a significant modification over the white-noise spectrum. Here, we use the fact that in the spherical collapse model, the fluctuations collapse and grow to size $M$ at redshift $z<z_{\rm eq}$ as isocurvature modes with $\delta(M) \propto a$ so that the transfer function can be approximated as a linear scale factor. Finally, we assume a Gaussian kernel function, whose Fourier transform is again a Gaussian function in $k=\lvert{\bf k}\rvert$ of the form
\begin{equation}
	W({\bf k},R) = \exp\left(-k^2R^2/2\right)\,.
\end{equation}
With this choice, the mass of a structure today that extends to a region of size $R$ is $M = (2 \pi)^{3/2} \rho_{\rm DM} R^3$.

With this parameterization, the standard deviation in Eq.~\eqref{eq:varianceR} is approximated by the analytic function
\begin{equation}
	\label{eq:varianceR1}
    \sigma(z, R) = \frac{1 + z_{\rm eq}}{1 + z} \sqrt{\frac{3}{5} \frac{\sqrt{\pi}\, {\rm erf}(x) - 2 x\, e^{-x^2}}{x^3}}\,,
\end{equation}
where $x = k_{\rm osc}R$ and ${\rm erf}(x)$ is the error function. Collapse occurs when $\sigma(z, R) \geq \delta_c$. More generally, the mass of the heaviest objects that form at redshift $z$, $M_{\rm max}(z)$, is found implicitly from the expression
\begin{equation}
	\sigma(z, M_{\rm max}(z)) = \delta_c\,,
\end{equation}
where $\sigma^2(z, M)$ is the variance corresponding to Eq.~\eqref{eq:varianceR} once $R$ is expressed in terms of $M$. The results for $z=0$ are reported in Fig.~\ref{fig:Minicluster} (blue line), where the kink at $m_a \sim 10^{-5}\,$eV corresponds to the effects of the kinetic term in the initial conditions. Whenever there exists a mechanism that grants a large initial kinetic energy for the axion field, the field would begin coherent oscillations in a colder universe, allowing for heavier axion miniclusters at MRE. The clumping of such heavier building blocks also leads to an increased value of the maximal mass, with observational consequences. Note, however, that these large miniclusters would probably not survive tidal stripping from other astrophysical objects such as brown dwarfs and main sequence stars, especially in high-density regions such as the Galactic center~\cite{Tinyakov:2015cgg, Kavanagh:2020gcy, Edwards:2020afl}.

\subsection{Stripping}
\label{sec:stripping}
It is not guaranteed that AMCs survive tidal stripping from compact objects in galaxies, such as brown dwarfs and stars. In all DM models, tidal interactions destroy small-scale clumps~\cite{Berezinsky:2003vn, Berezinsky:2014wya}, such as axion miniclusters~\cite{Tinyakov:2015cgg}. Recent N-body simulations have proven that the stripping mechanism is crucial for the population of miniclusters in galaxies~\cite{Kavanagh:2020gcy}. In general, it could be expected that the larger and heavier miniclusters produced in the strong KM regime would be more prone to get tidally disrupted by compact objects.

The effect of the encounter of the minicluster with an individual compact object of mass $M$ with relative velocity $v_{\rm rel}$ is that of increasing the velocity dispersion of the bounded axions. An encounter that occurs close enough would deposit sufficient energy so that the minicluster is completely disrupted. This occurs for an impact parameter $b$ smaller than the critical value~\cite{Goerdt:2006hp, Schneider:2010jr}
\begin{equation}
    b < b_c \equiv \left(\frac{G M R_{\rm AMC}}{v_{\rm rel}\,v_{\rm AMC}}\right)^{1/2},
\end{equation}
where the velocity dispersion of the minicluster is $v_{\rm AMC}^2 = G\, M_{\rm AMC}/R_{\rm AMC}$. The probability of disruption for a minicluster moving in a stellar field of column mass density $S$ is~\cite{Tinyakov:2015cgg}
\begin{equation}
    \label{eq:survival}
    p_{\rm disr} = 2\pi\, S\,\frac{G\, R_{\rm AMC}}{v_{\rm rel}\, v_{\rm AMC}}\,.
\end{equation}
In the vicinity of the solar system, it is generally found $p_{\rm disr} = \mathcal{O}\left(10^{-2}\right)$ and miniclusters generally survive the stripping process. However, this result depends on the density of the minicluster and not on its mass or radius separately. Since the density of miniclusters given in Eq.~\eqref{eq:rhomc} is related to the spherical collapse model and not to the cosmological history, we generally expect that at the lowest order in which this approximation holds, the probability of disruption would not change among the different scenarios of misalignment mechanisms. Near the Solar system, Eq.~\eqref{eq:survival} yields $p_{\rm disr} \approx 2\%$.

\subsection{Microlensing}
\label{sec:microlensing}

Now we discuss a possible method to distinguish between the different misalignment scenarios using microlensing. Bound structures made of axions such as miniclusters, minicluster halos, and axion stars can impact lensing from distant sources. For a point-like lens of mass $M$, the characteristics of the microlensing event are determined by the Einstein radius~\cite{Paczynski:1985jf, Griest:1990vu}
\begin{equation}
    \label{eq:einsteinRad}
    R_E(x) = \sqrt{\frac{4\,G_{N}\,M}{c^2}\frac{D_{L}\,D_{LS}}{D_{S}}} \approx  4.3\! \times \!10^3{\rm \, km}  \left[\!x(1\!-\!x) \frac{D_{S}}{\rm kpc}\frac{M}{10^{-10}M_{\odot}}\!\right]^{\!1/2}\!\!,
\end{equation}
where $D_{S}$, $D_{L}$, and $D_{LS}$ are the distances between the source and the observer, the lens and the observer, and the source and the lens, respectively, and $x = D_L/D_S$.

Although lensing can occur from axion miniclusters and halos~\cite{Kolb:1995bu, Fairbairn:2017dmf, Fairbairn:2017sil, Kavanagh:2020gcy}, these structures can generally not be modeled as point lenses, as their Einstein radius lies within their mass distribution, so that the internal density profile must be known to estimate the lensing power. Extended objects generally lead to weaker limits due to the smaller magnification of the lens~\cite{Croon:2020ouk, Croon:2020wpr}.

Here, we focus on the lensing of light coming from a distant source when the lensing object is an axion star, which is generally a much more compact object than a minicluster~\cite{Barranco:2010ib, Chavanis:2011zi, Eby:2014fya, Visinelli:2017ooc, Schiappacasse:2017ham}. Axion stars belong to the class of real scalar field oscillatons~\cite{Seidel:1991zh, Copeland:1995fq, Urena-Lopez:2001zjo, Visinelli:2021uve} in which the field occupies the lowest energy state that is allowed by the Heisenberg uncertainty principle. An axion star of mass $M_{\rm as}$ is generally produced in the dense core of axion miniclusters through the mechanism of gravitational cooling~\cite{Seidel:1993zk} with the relaxation time $\tau_{\rm as}$~\cite{Levkov:2018kau}. Although the decay rate in two photons does not significantly affect the stability of axion stars on a cosmological time scale, self-interactions of type $3\,a\to a$ can lead to the decay of axion stars with a decay rate that depends on $\left(m_a/f_a\right)^2$~\cite{Eby:2015hyx}.

An axion star can efficiently lens the light from a distant source, as their radius is typically smaller than their corresponding Einstein radius. Recently, a single microlensing event observed by the Subaru Hyper Suprime-Cam (HSC) collaboration after observing in M31 ($D_S = 770\,$kpc) for 7 hours~\cite{Niikura:2017zjd} has been interpreted in terms of an axion star of planetary mass~\cite{Sugiyama:2021xqg} (see also Ref.~\cite{Schiappacasse:2021zlr}). However, the result is of difficult interpretation in the standard scenario of the misalignment mechanism, since such massive stars would only form for the lightest QCD axions, for which the post-inflation scenario does not hold. Although the AMCs formed in the strong KM regime are generally much heavier, the axion stars formed within them are not expected to differ much from those of the standard scenario, since the mass of the axion star is only mildly dependent on that of the host AMC as $M_{\rm as} \propto M_{\rm AMC}^{1/3}$~\cite{Schive:2014hza}. For this reason, the KM regime cannot be invoked to produce the heavy axion stars which are needed to explain the microlensing results in Ref.~\cite{Sugiyama:2021xqg}. One issue with the derivation of the axion star properties is that the formulas used rely on the results of Ref.~\cite{Schive:2014hza} which are obtained for ultralight axions for which the mass scale greatly differs from that of the QCD axion. Although this can be justified since the set of equations describing the system (the Newton-Poisson equation) possesses a scaling property, a dedicated simulation proving this is yet lacking.

\section{Conclusions}
\label{sec:concl}

An explicit breaking term in the Peccei-Quinn (PQ) symmetry could give rise to a non-zero velocity term for the axion field. This scenario, called kinetic misalignment (KM), has been explored in the literature in relation with baryogenesis. 
Even if the models of KM presented in the literature typically correspond to a PQ symmetry breaking happening during inflation, in this work we focused on the low-energy dynamics of an axion featuring a non-zero velocity, being agnostic about how it was produced in the early Universe.

Here, we have discussed further implications for the delayed onset of the oscillations in the axion field that appear in the KM scenario. In the standard scenario, the DM axion mass depends on the relative size of the inflation scale with respect to the axion energy scale: in the pre-inflation regime, the initial misalignment angle can be tuned to achieve a specific mass scale according to the result in Fig.~\ref{fig:Oh2_th}, while in the post-inflation regime the mass is fixed by the dynamics of the axion field that yields $\theta_i = \mathcal{O}(1)$. In KM scenarios, different values of the DM axion mass can be explored because of the presence of the non-zero initial velocity as a new parameter upon fixing $\theta_i$, see Fig.~\ref{fig:Y-ma}.

One aspect that has been explored here is the formation of axion miniclusters (AMCs) and minihalos in KM regimes, as a possible tool that leads to distinctive signatures from the standard scenario. AMCs are generally formed in the post-inflation regime with typical mass $M_0 \sim 10^{-11}~M_\odot$. In KM scenarios, the non-zero velocity term allows for a wider mass range for the AMCs: the mass of the AMCs is larger than what is obtained in the standard scenario because the axion field begins to oscillate in a colder universe with a larger comoving scale, as shown in Fig.~\ref{fig:Minicluster}. In this regime, AMCs are more diffuse and heavier, so that assuming that the fraction of axions in bounded structures is the same, there would be fewer of them and they would be affected by tidal stripping the same way as in the standard scenario. The clumping of heavier AMCs would lead to larger halos of miniclusters, with the typical mass today that could be orders of magnitude above what has been expected so far, and could affect the analysis of the microlensing events from minicluster halos.

Future directions would involve employing a numerical solution of the equation of motion, including the effects of explicit symmetry breaking. For example, the dynamics of the axion can be resolved by modifying the recent open-source numerical routine \texttt{MiMeS}~\cite{Karamitros:2021nxi}. The properties of AMCs can only be assessed through more sophisticated analyzes that account for the evolution of the PQ field and require modification of existing open-source codes that are already available~\cite{Vaquero:2018tib}. A similar analysis involving the implementation of a Boltzmann solver can be performed in the pre-inflationary scenario, where the KM regime would be additionally constrained by isocurvature fluctuations, as has been discussed in Ref.~\cite{Co:2020dya}.

\section*{Acknowledgements}
NB thanks Nicol{\'a}s Fern{\'a}ndez for fruitful discussions. BB and NB received funding from the Patrimonio Aut{\'o}nomo - Fondo Nacional de Financiamiento para la Ciencia, la Tecnolog{\'i}a y la Innovaci{\'o}n Francisco Jos{\'e} de Caldas (MinCiencias - Colombia) grant No.~80740-465-2020. NB is also funded by the Spanish FEDER/MCIU-AEI under grant FPA2017-84543-P. NR thanks Pedro Schwaller, Enrico Morgante \& JGU THEP Group for great discussion and acknowledges support by the Cluster of Excellence ``Precision Physics, Fundamental Interactions, and Structure of Matter'' (PRISMA+EXC 2118/1) funded by the German Research Foundation (DFG) within the German Excellence Strategy (Project ID 39083149). LV acknowledges support from the European Union's Horizon 2020 research and innovation programme under the Marie Sk{\l}odowska-Curie grant agreement No.~754496 (H2020-MSCA-COFUND-2016 FELLINI). This project has received funding and support from the European Union's Horizon 2020 research and innovation programme under the Marie Sk{\l}odowska-Curie grant agreement No.~860881 (H2020-MSCA-ITN-2019 HIDDeN). 

\authorcontributions{All authors have contributed equally to the preparation of the manuscript.}

\funding{This research received no external funding.}

\institutionalreview{Not applicable.}

\informedconsent{Not applicable.}

\dataavailability{Not applicable.} 

\conflictsofinterest{The authors declare no conflicts of interest.} 

\bibliography{biblio}

\end{document}